\documentclass{article}

\usepackage{arxiv}
\pdfoutput=1

\usepackage[utf8]{inputenc} 
\usepackage[T1]{fontenc}    
\usepackage{hyperref}       
\usepackage{url}            
\usepackage{booktabs}       
\usepackage{amsfonts}       
\usepackage{nicefrac}       
\usepackage{microtype}      
\usepackage{graphicx}

\usepackage{amsmath}
\usepackage{graphicx}
\usepackage{multicol}
\usepackage{multirow}
\usepackage{subcaption}
\usepackage{caption}
\usepackage{hyperref}
\usepackage{makecell}
\usepackage{diagbox}

\title{weg2vec: Event embedding for temporal networks}

\author{
  Maddalena ~Torricelli\\
  ISI Foundation, Italy\\
  Alma Mater Studiorum University of Bologna, Italy\\
  \texttt{maddalena.torricell2@unibo.it} \\
   \And
 M\'arton ~Karsai \\
  Department of Network and Data Science, Central European University, H-1051 Budapest, Hungary\\
  Univ de Lyon, ENS de Lyon, INRIA, CNRS, UMR 5668, IXXI, 69364 Lyon, France\\
  ISI Foundation, Italy\\
  \AND
 Laetitia ~Gauvin \\
  ISI Foundation, Italy\\
}

\begin{document}
\maketitle

\bigskip
\begin{abstract}
Network embedding techniques are powerful to capture structural regularities in networks and to identify similarities between their local fabrics. However, conventional network embedding models are developed for static structures, commonly consider nodes only and they are seriously challenged when the network is varying in time. Temporal networks may provide an advantage in the description of real systems, but they code more complex information, which could be effectively represented only by a handful of methods so far. Here, we propose a new method of event embedding of temporal networks, called \emph{weg2vec}, which builds on temporal and structural similarities of events to learn a low dimensional representation of a temporal network. This projection successfully captures latent structures and similarities between events involving different nodes at different times and provides ways to predict the final outcome of spreading processes unfolding on the temporal structure.
\end{abstract}

\section{Introduction}
An interacting group of people, the collectively active neurons in the brain, or the transportation system of a city are only a few examples of complex systems, which are all intrinsically dynamical~\cite{bar1997dynamics}. They can be commonly interpreted as a set of entities, which interact over time and form a network structure coding the architecture of the system in hand~\cite{newman2003structure,albert2002statistical}. This duality of the structure and temporal nature of interactions can be effectively captured by temporal networks, proposing a new and precise representation of complex systems as compared to earlier strategies~\cite{holme2012temporal,masuda2016guide}. On the finest level, temporal networks consist of time-varying events between interacting nodes, and as a whole they appear as systems with high complexity and dimensionality. Events in real temporal structures, however, are not random but correlated with each other and arguably driven by several microscopic mechanisms leading to several generative characters of the network. Emerging properties like the heterogeneous number or strength of interactions~\cite{perra2012activity,karsai2014time}, community structure~\cite{laurent2015calls,gauvin2014detecting,peixoto2017modelling}, degree correlations, bursty temporal patterns~\cite{barabasi2005origin,karsai2018bursty}, or temporal motifs of causally correlated events~\cite{kovanen2011temporal,paranjape2017motifs} are all arguably induced by such mechanisms. In turn, temporal events and their correlations largely influence dynamical processes as well~\cite{masuda2017temporal}, like they determine the speed and final outcome of information or epidemic spreading~\cite{karsai2011small,gauvin2013activity,delvenne2015diffusion,scholtes2014causality}. The recognition of these impacts has put temporal networks in the focus point of several investigations recently, which yet struggle to propose efficient representations, which capture the complex temporal/structural information coded in them, while reducing their dimensionality to ease their analysis.

Correlated patterns in the structure and dynamics of networks usually can be captured by certain higher-order representations~\cite{benson2016higher}. For static networks, \emph{line graphs} propose an efficient description~\cite{evans2009line,ahn2010link}, which in their simplest form identify static links as nodes and connect them if they are adjacent, i.e., share an ending node in common. Other technique is based on simplicial complexes~\cite{petri2013topological} considering homology of the network topology to capture higher order structures. At the same time, recent \emph{network embedding} methods propose inventive ways to obtain a reduced dimensional representation of static structures. Their common goal is to map complex networks to a low-dimensional space, while conserving certain similarities of nodes reflected by some distance metrics in the embedding. Most common methods use random walk sampling~\cite{perozzi2014deepwalk,grover2016node2vec} or graph convolution~\cite{hamilton2017inductive,kipf2016semi} to capture the local structural context of network nodes. In case of temporal networks, the recently proposed \emph{event graph} representation~\cite{kivela2017mapping,mellor2017temporal} defines a higher-order description by identifying relations between events, which are adjacent, i.e. not simultaneous and share at least one ending node. Adjacent events then are connected by links with direction respecting the arrow of time, and with weight defined as the absolute time difference between the connected events. This way of description is very useful as it codes all time-respecting paths in a temporal network at once, while proposing an information lossless representation of the temporal structure as a static weighted directed acyclic graph. Recently a few \emph{dynamical network embedding} methods have been developed to consider dynamical changes in the structure in the learned network representations. At the base of many methods there is the modification of the standard representation of the temporal network, whether it is in the form of a list of events, a tensor \cite{gauvin2014detecting} or a supra-adjacency matrix \cite{2019arXiv190905976S}. All of these methods~\cite{pandhre2018stwalk,singer2019node,beres2018node} commonly aim to solve a node embedding problem by locally sampling the temporal-structural neighbourhood of nodes to create contexts, which they feed to a Skip-Gram learning architecture borrowed from the text representation literature~\cite{mikolov2013efficient}. As a solution, they build a sequence of correlated/updated embeddings of network snapshots, which consider short term history of the network backward in time. However, these methods have some common limitations; first of all, it can be hard to manage a high number of hyper parameters for the control of the sampling random walk process and the embedding itself. At the same time, the embedding of nodes may miss to reflect the dynamical changes of temporal interactions. Finally, taking into account only past and present interactions in the embedding can crucially limit the performance of the prediction, while the consideration of future events can significantly improve this task.

Here we propose a new temporal network embedding method, we call \emph{weg2vec} (weighted event graph 2 vector), which aims to tackle all these shortcomings. This is an event embedding method, which represents an entire temporal network in the same reduced dimensional abstract space. It is based on combined event contexts built by sampling locally a higher-order static representation of temporal networks, which in turn code the complex patterns characterising the structure and dynamics of real world networks. This is an unsupervised representation learning technique, which can consider the past and future context of an event simultaneously. It is sampling without using dynamical processes, thus it can be controlled by a handful of hyper parameters. It identifies similarity between different events/nodes, which may be active at different times, but influence a similar set of nodes in the future. To demonstrate the power of this representation, we showcase its utility via predicting the final outcome of modelled spreading processes on several real world temporal networks. This prediction task performs significantly better when it builds on our representation as compared to other dynamical network embeddings.

As follows, first we will present the pipeline to build our embedding method. We will show the characteristics of our representation, measuring its stability and its ability to capture temporal and structural information from the network. We will show then the results obtained in estimating an epidemic spreading outcome, as a showcase of the potentials of our embedding method in the analysis of dynamic processes. Finally, we will compare our results to similar computations performed with two other embedding methods. In the final Sections, we present the discussion of the results and the analysis of the methods.

\section{Results}
\label{sec:methods}
\begin{figure}[h!]
\centering
\includegraphics[scale=0.65,trim={0.02cm 4.5cm 0.04cm 3.2cm},clip]{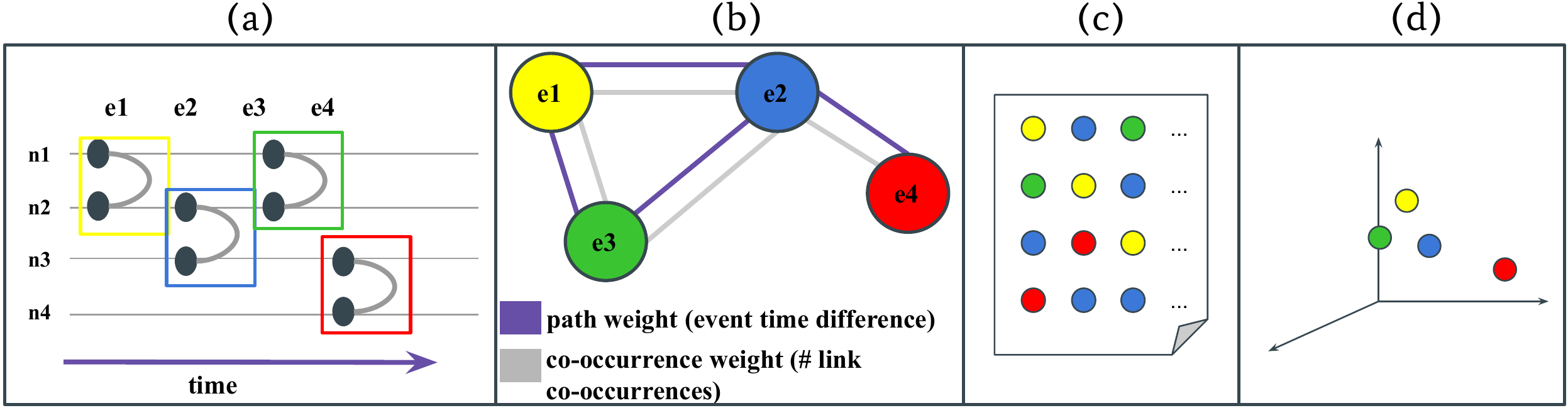}
\caption{Schematic presentation of the methodological pipeline of the presented temporal network embedding method, which takes a temporal network to (a) project it  into a weighted event graph; (b) to sample a set of environments for each event; (c) and uses it as input for a Skip-Gram model; (d) to obtain an event embedding of the original network.}
\label{fig:pipeline_visualization}
\end{figure}

An embedding method of temporal networks may take a list of temporal interactions as input, and provide a lower dimensional representation, in which vectors corresponding to similar nodes or events in the original structure ideally point close to each other in the embedding. In our pipeline we solve this problem in three consecutive methodological steps. First, we turn the original temporal network into a higher-order representation, which captures pairs of adjacent and consecutive events, which may be in causal relationship. Second, we use this representation to generate environments for each event, sampled from their adjacent neighbours. Finally we obtain an embedding of events by training a Skip-Gram model on the obtained environments. These steps are schematically drawn in Figure \ref{fig:pipeline_visualization}) and introduce next in the coming sections.

To demonstrate our method we used four different datasets all obtained from the SocioPatterns project~\cite{cattuto2010dynamics}. These open datasets record the time evolving physical proxy interactions of a large number of people in different settings like in \emph{conference}, \emph{high school}, \emph{hospital}, or \emph{primary school}. The data comes as a long sequence of network snapshots recording simultaneous interactions between any participants in every $20$ seconds. While for demonstration most of the result in the paper are shown only for the conference and primary school settings, a detailed data description together with the analysis for the other networks are presented in the Supplementary Information (SI).

\subsection{Temporal networks as weighted event graphs}
\label{sec:tn_dag}

Let us consider a temporal network
\begin{equation}
G_T=(N,E_T,T),
\end{equation}
where $E_T$ denotes a set of events (temporal edges) among nodes in $N$ at times $t \in T$. Specifically, we define an event $e=(i,j,t)$ as an interaction between two nodes $(i,j) \in N \times N$ at a given timestamp $t \in T$. The time aggregation of interactions in $G_T$ over $T$ maps the underlying structure into a static graph $G=(N,E)$ defined over the same set of nodes $N$, which are pairwise connected by links $(i,j)\in E$ if at least once interacted. For simplicity here we assume that events are undirected and no self events/links exist, i.e. for any event $(i,j,t)$ or link $(i,j)$, $i \neq j$.

We define two events $e_1=(i,j,t_1)$ and $e_2=(j,k,t_2)$ to be \emph{adjacent} ($e_1 \rightarrow e_2$) if they share at least one node ($\{i,j\}\cap \{j,k\}\neq \emptyset$) and they are time consecutive ($t_1<t_2$)\footnote{Note that adjacency in a static network $G$ can be similarly defined between links $l_1=(i,j)$ and $l_2=(j,k)$, which share at least one node ($\{i,j\}\cap \{j,k\}\neq \emptyset$).}. More restrictively, we called them $\delta t$-\emph{adjacent} ($e_1 \xrightarrow{\delta t} e_2$) if they are adjacent and $|t_2-t_1|\leq \delta t$, thus follow each other within a given period of time ($0\leq \delta t \leq T$). Adjacency introduces a directed relation between events, with orientation respecting their timely order. Using this notion we can formally define a static directed network representation $D=(E_T,E_D)$ of any temporal network, where original events in $E_T$ are defined as nodes and they are connected by directed links $e_{D}\in E_D$ if they are adjacent $e_D=e_1 \rightarrow e_2$. The obtained network is a directed acyclic graph called the \emph{event graph}, defined earlier in~\cite{kivela2018mapping,mellor2017temporal}. It can be interpreted as a temporal line graph, which provides a higher-order representation to map out simultaneously all time respecting paths of the original temporal network without any loss of information. Note, that to simplify our representation, for a given event if it has multiple future adjacent events with the same pair of nodes, we only consider the earliest one.

Event graphs can be easily enriched with various types of link weights reflecting some temporal or structural information coded in the original structure. Here, to better capture the strength of potential causal relationships, first we consider a weight defined as $w_{path}=\frac{1}{(1+|t_2-t_1|)}$, which is a measure inversely proportional to the absolute time difference between adjacent events at $t_1$ and $t_2$. At the same time we define a second weight for adjacent events (links of the event graph), based on the total number of co-occurring events on the underlying adjacent links in the static network. More precisely, the $w_{co}(e_1,e_2)$ co-occurrence weight counts the number of $\delta t$-adjacent events in $G_T$ appearing on a given pair of adjacent links $l_1(i,j)$ and $l_2(j,k)$ in the static graph $G$. Note that adjacent events connected in $D$, which corresponds to the same links in the underlying network $G$, will have the same $w_{co}$ values\footnote{Datasets analysed in this paper are defined as sequences of snapshots aggregating temporal interactions in consecutive time windows of size $\Delta t$. In these systems we compute $w_{co}$ for adjacent links as the number of co-occurrence of corresponding events within single snapshots. This definition may slightly underestimate the real co-occurrence (if $\delta t\leq\Delta t$), but provides the best plausible solution due to the un-ambiguity of timings of events within a single snapshot.}.

\subsection{Neighbourhood sampling strategy}
\label{sec:context}

In the same spirit of recent node embedding techniques~\cite{mikolov2013efficient, grover2016node2vec} based on the Skip-Gram model, we propose an event embedding method, which samples neighbourhoods for events from the weighted event graph representation to map them to a lower dimensional space. To assign an environment to an event $e_k$, we sample its local neighbourhood set $N_k$, which consists of the set of its first in- (past) and out- (future) neighbours (also called its \textit{predecessors} and \textit{successors} from now on). The sampling is done according to probabilities determined by the two types of weights of the links that connect the actual event to its neighbours.

In order to consider not only the past but the future of an event in its environment, during sampling we use a combined set of its predecessor and successor events. Further, to balance the contribution of causality and temporal co-occurrence, we introduce a neighbourhood sampling strategy such that the probability $p(e_l)$ of picking an event $e_l$ from the combined neighbourhood set $N_k$ of the central event $e_k$ is given by :
\begin{equation}\label{eq:proba}
p(e_l) = \alpha F(w_{path}(e_k, e_l))  + (1 -\alpha) F(w_{co}(e_k, e_l))
\end{equation}
where $\alpha$ is a coefficient between $0$ and $1$ scaling the contribution of the two types of weights and $F$ is a normalised weighted function defined as:
\begin{equation}\label{eq:proba_F}
\centering
F(w) = \frac{w(e_k, e_l)}{\sum\limits_{n \in N_k}^{} w(e_k, e_n)} \cdot
\end{equation}
Using such probabilities computed for each neighbour, we sample $nb$ number of environments randomly for each event with length $s$ for an effective training of the model explained next.

\begin{figure}[h!]
\centering
\includegraphics[width=0.9\linewidth]{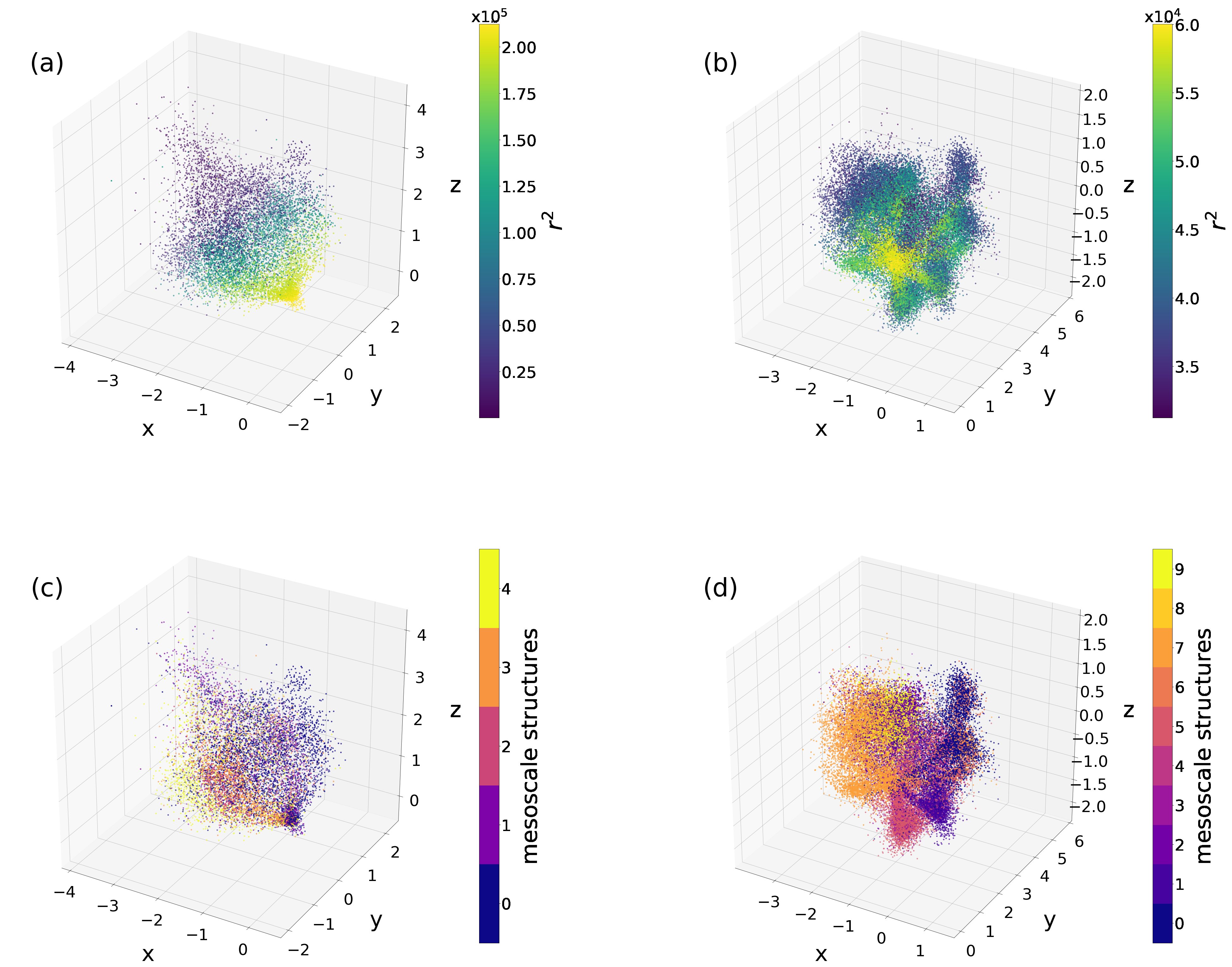}
\caption{3-dimensional embeddings of the conference and primary school networks. $x$,$y$ and $z$ axes indicate event coordinates, while colour in panels (a-b) shows the time at which the event occurs and in panels (c-d) mesoscale structure membership found using a tensor factorisation method (see Section \ref{sec:tensor_decomposition}) respectively set to find $5$ and $10$ of these structures. Hyper parameters were set to $\alpha=0.5$, $nb=10$ and $s=10$.}
\label{fig:emb3d}
\end{figure}

\subsection{Embedding of empirical temporal networks}
\label{sec:embedding}

Once the environments have been created, they are passed as inputs to the Skip-Gram model, with parameters fixed to different values according to the analysis to conduct. The result is a $d$-dimensional vector space in which events are represented by Cartesian coordinates. As an illustration, Figure~\ref{fig:emb3d} shows a $3$-dimensional embedded representation of two of the empirical networks we used for the analysis, recorded in a conference and a primary school settings. Our aim with this setting was to investigate the performance of low-dimensional embedding on the one hand, and on the other taking into account equally the effects of causal temporal paths and co-occurrences by setting $\alpha=0.5$. The environment parameters $nb$ and $s$ have been set both to $10$ as an example. In Figure~\ref{fig:emb3d} (a) and (b) each event is represented as a point in the embedded space with colour indicating the time at which they occurred in the original temporal network. Interestingly, the gradient change of colours indicates that these embeddings capture in large part the time ordering of the events. At the same time, Figure~\ref{fig:emb3d} (c) and (d) shows the same $3$-dimensional embedded representations, but with colours representing the membership to mesoscale structures detected by tensor factorisation methods applied on the original temporal network \cite{gauvin2014detecting} (see Section on \ref{sec:tensor_decomposition}). Evidently, as colours are not distributed randomly but similar colours are somewhat grouped together in space, it suggests that our embedding is able to capture some of these mesoscale structures as well.

\subsection{Effects of the dimension and of the neighbourhood sampling}
\label{sec:entropy}

One of the most important hyper parameter of our method is the number of dimensions of the embedding vector space. Lower than optimal number of dimensions may lead to neglected but otherwise relevant latent correlations in the temporal structure, while overestimation of this number may give us a highly redundant embedded space. We test here the consistency and robustness of our embedding technique in terms of this parameter. We argue that as we increase the number of dimensions, once it reaches and overpasses an optimal number, it starts coding increasingly redundant information in the embedding. As a consequence, further dimensions would not alter the relative positions of embedded events and the pairwise euclidean distances among them stabilise. To check this assumption, we use an entropy measure capturing the fluctuations of pairwise euclidean distances over several realisations with the same number of embedding dimensions. For selected event pairs, we measure the probability distribution of their pairwise euclidean distances over $10$ realisations (see Section \ref{sec:entropy_method}) and use it to compute an entropy score. Averaging these scores over all the event pairs provides us an indicator of the stability of the embedding.

\begin{figure}[h!]
\centering
\includegraphics[width=0.9\linewidth]{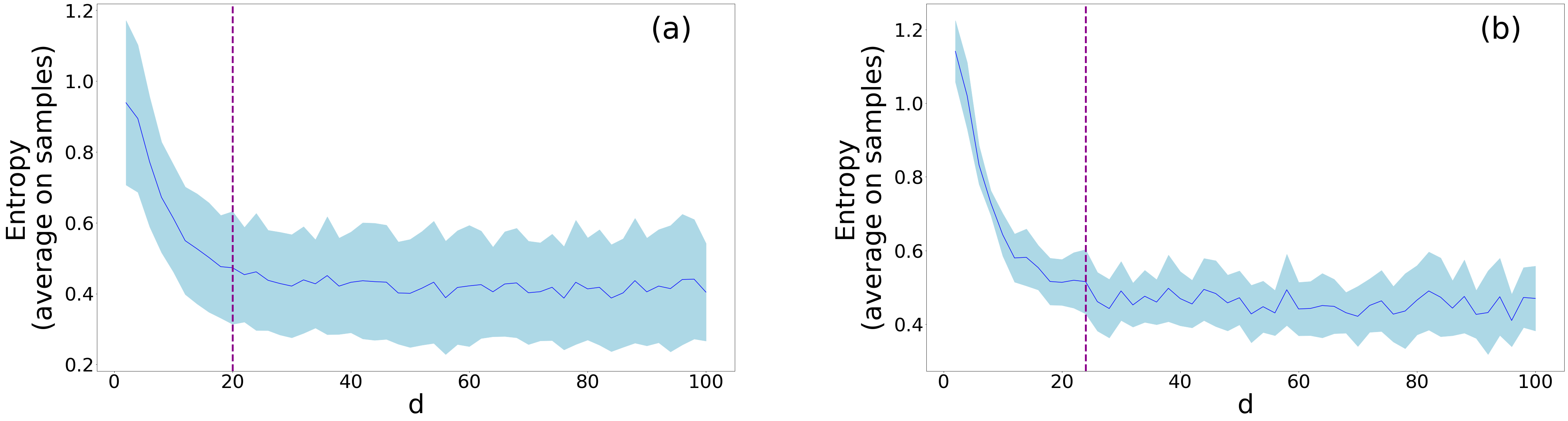}
\caption{Entropy values with respect to $d$ number of embedding dimensions for the conference (a) and the primary school (b) networks at $\alpha = 0.5$. The dash line represents the value ($d=20$ and $24$ respectively for (a) and (b)) of the optimal embedding dimension in which stability is reached. The blue line and the shaded area represent respectively the average and the variance among the samples we used for the analysis.}
\label{fig:entropy_01}
\end{figure}

We show results in Fig.~\ref{fig:entropy_01} for two empirical networks using a balanced embedding with $\alpha=0.5$ in both cases. As we expected, the entropy decreases as we increase the number of dimensions. This is due to stabilisation of the distribution of pairwise euclidean distances, which in turn gives us a hint on the optimal dimension at which the local neighbourhoods (as defined by the sampling) are well captured by the embedding. To select this optimal dimension, we identified the lower bound at which entropy starts to fluctuate around a constant value (see Methods \ref{sec:entropy_method} and SI). 

\subsection{Spreading process prediction with embedding events}
\label{sec:validation}

Beyond the demonstrated capacities of our model to capture the temporal ordering and the underlying mesoscopic structures, it may provide further useful information about the embedded events. As a temporal network embedding, it positions events in proximity with similar neighbourhoods. In other words, it can help to identify similar events maybe involving different nodes at different times, but influencing a similar set of other nodes via overlapping temporal paths. As a consequence, this information can be used to predict the outcome of information diffusion processes on temporal networks.

To explore this problem, we model a Susceptible-Infected (SI) process, which is the simplest schematic model of epidemic or information spreading (see Section \ref{sec:spreading} in Methods). Defined on networks, this model assumes that each node can be in one of two mutually exclusive states (susceptible (S) or infected (I)) at a given time. While initially each node is susceptible, infection can spread from a selected infected seed node/event via temporal interactions and can reach all other nodes via connected valid temporal paths. To obtain the expected outcome of SI process on a temporal network we took each event as the seed and simulated the spreading on the empirical temporal network to measure the final epidemic size in each case. To test the versatility of our embedding method we trained a model using the embedded coordinates of events for epidemic size predictions and compared results directly to the corresponding simulated outcomes. We used linear regression to approximate the correspondence between the embedding coordinates of each event and the size of epidemic initiated from them. As the goodness of the prediction we simply computed the $r^2$ scores between the predicted and simulated epidemic sizes. Note, that we tested more complicated non-linear models but obtained lower performance in prediction (not shown here). We report our results in Table \ref{tab:general_results}, where we fixed the environment parameters $s$ and $nb$ both to $10$ and chose the optimal embedding dimension for each real network detected as we explained in Section~\ref{sec:entropy} and the SI.

Real temporal networks are interwoven by several temporal and structural correlations. First there are \emph{local temporal correlations} induced by intrinsic or environment effects emerging between events on the same link leading to bursty behaviour, event trains or circadian activity patterns. Another type is \emph{higher-order temporal correlations} leading to the emergence of causal temporal motifs. \emph{Structural correlations} are responsible for the emerging assortative patterns, communities, or any non-random connection pattern in a social network, while \emph{weight-structural correlations} induce the non-random distribution of strong and weak ties inside and between communities. In Table~\ref{tab:RRMcorr} we summarise which RRM preserves and eliminates which type of correlations.

To get a better sense about the effects of these correlations, we used three types of randomised reference models (RRM) \cite{gauvin2018randomized} to eliminate combinations of temporal and structural correlations, and to identify which of them are determinant for the prediction task. These RRMs were
\begin{itemize}
\item \textit{Snapshot shuffling}, which randomises the timestamps of events in order to eliminate any temporal correlation between them inducing burstiness, causal motifs, or group activation etc. This model is assigned as P[$\textbf{f}(\textbf{t}),p_{T}(\Gamma)$] using the notation convention introduced in~\cite{gauvin2018randomized}.
\item \textit{Timeline shuffling} takes the complete timeline of events between a connected pair of nodes in the temporal network and swap it with the timeline of another randomly selected connected pair of nodes. This shuffling method, noted as P[$L,\textbf{f}(p_{L}(\Theta))$] in~\cite{gauvin2018randomized}, eliminates all correlations between the underlying structure and timelines while also vanish any casual correlations between events on adjacent links.
\item \textit{Link shuffling} method (noted as P[$\textbf{f}(L),p_{L}(\Theta)$] in~\cite{gauvin2018randomized}) randomises links of the underlying aggregated static network first to obtain a Bernoulli random structure, and then reassign randomly the original timelines of events to the new links without replacement. In this way, it destroys any structural and structural-temporal correlations in the network, while keeping local temporal correlations like burstiness unaltered.

For a summary of present and eliminated correlations in the different RRMs see Table~\ref{tab:RRMcorr}.

\end{itemize}
\begin{table}[ht]
\begin{center}
\begin{tabular}{|l l||c|c|c|c|c|}
\hline
\multicolumn{2}{|l|}{\backslashbox{\textbf{RRM}}{\textbf{Correlation}}}
 & $\vcenter{\hbox{Local}\hbox{temporal}}$ & $\vcenter{\hbox{Weight-}\hbox{structural}}$ & $\vcenter{\hbox{Higher-order}\hbox{temporal}}$ & Structural \\ \hline\hline
 Original & & \checkmark & \checkmark & \checkmark & \checkmark \\ \hline
 Snapshot & P[$\textbf{f}(\textbf{t}),p_{T}(\Gamma)$] & $\times$ & \checkmark & $\times$ & \checkmark \\ \hline
 Timeline & P[$L,\textbf{f}(p_{L}(\Theta))$] & \checkmark & $\times$ & $\times$ & \checkmark \\ \hline
 Links  & P[$\textbf{f}(L),p_{L}(\Theta)$] & \checkmark & $\times$ & $\times$ & $\times$ \\
\hline
\end{tabular}
\caption{Summary of preserved and eliminated structural and temporal correlations (Local temporal, Weight-structural, Higher-order temporal, and Structural) in the original and different random reference models (Snapshot, Timeline and Link shuffling) of temporal networks. For further explanation see text.}
\label{tab:RRMcorr}
\end{center}
\vspace{-5mm}
\end{table}

The prediction results for the original and the RRM networks are summarised in Table \ref{tab:general_results}, where we depict the observed average $r^2$ values with their standard deviation computed over the embedding realisations. These results suggest that in general the randomised model embeddings perform worse in predicting the final epidemic size with respect to the original network embeddings. In one way this is straightforward as some correlations have been eliminated from RRMs, which might be determinant for the prediction task. On the other hand, RRMs also appear with a less complex structure and limited irrelevant dependencies and noise, which in turn may help the prediction. It is the snapshot shuffling method, which leads consistently to a significant drop in performance, suggesting that temporal correlations (local or higher-order) are very important for the spreading process and that our embedding can capture these dependencies successfully. Timeline shuffling, which destroys weight-structural and higher-order temporal correlations but conserve the dynamics on links and the underlying network seems to perform better as compared to the snapshot shuffling method but yet worse than the original network. This suggests that while structural correlations can be captured by the embedding, local temporal correlations might be better predictors than weight-structural correlations. Interestingly, the link shuffling method, which conserves only local timeline dynamics on links, performs the best among the RRMs, sometimes even better than the original dataset. Consequently, indeed local temporal event dynamics is the most important feature of the temporal network, while removing structural correlations the system becomes homogeneous and easier to predict (for a supporting analysis see SI). Although these general conclusions seem to be consistent over the several analysed datasets, results computed in different settings may have some fluctuations. As explained in the SI, this can be partially explained by the variance of the epidemic size distribution reflecting the fluctuation of epidemic size started from different times and events. In most of the cases smaller variance of epidemic size correlates with higher predictive performance except for the link shuffling method, as explained in the SI.

\begin{table}[ht]
\begin{center}
\begin{tabular}{|l l||c|c|c|c|}
\hline
\multicolumn{2}{|l||}{\backslashbox{\textbf{Data}}{\textbf{$r^2$}}}
 & \textbf{Original} 
 &\textbf{Snapshot} &\textbf{Timeline} &\textbf{Link}\\
\hline\hline
Conference & (d=20) &0.79$\pm$0.01 
& 0.53 $\pm$ 0.04  & 0.66 $\pm$ 0.03 & 0.57 $\pm$ 0.01 \\
\hline
Hospital & (d=14) &0.53$\pm$0.03 
& 0.11 $\pm$ 0.02  & 0.35 $\pm$ 0.06 & 0.50 $\pm$ 0.04\\
\hline
High School & (d=26) &0.56$\pm$0.02 
& 0.23 $\pm$ 0.01 & 0.53 $\pm$ 0.02 & 0.76 $\pm$ 0.04 \\
\hline
Primary School & (d=24) &0.68$\pm$0.02 
& 0.12 $\pm$ 0.01  & 0.31 $\pm$ 0.02 & 0.55 $\pm$ 0.02\\
\hline
\end{tabular}
\caption{R-squared values, $r^{2}$ obtained between the simulated and predicted epidemics outcomes using embedding of the real empirical temporal networks and of the randomised model. We set the environment parameters $s$ and $nb$ both to 10. The optimal embedding dimension were chosen as found in \ref{sec:entropy}.}
\label{tab:general_results}
\end{center}
\vspace{-5mm}
\end{table}

As a general conclusion we showed that the embedding successfully captures a combination of temporal and structural features of the network. On the other hand, the fact that temporal and structural features can be entangled has an impact on embedding performance but not on what the embedding is able to learn about them. For instance, the model can under-perform in a community-rich network where nodes of the same community have totally uncorrelated activities, while can provide precise predictions in shuffled datasets if structural and temporal correlations code redundant information.

\subsection{Parameter dependencies}

\begin{figure}[h!]
\centering
\includegraphics[width=0.9\linewidth]{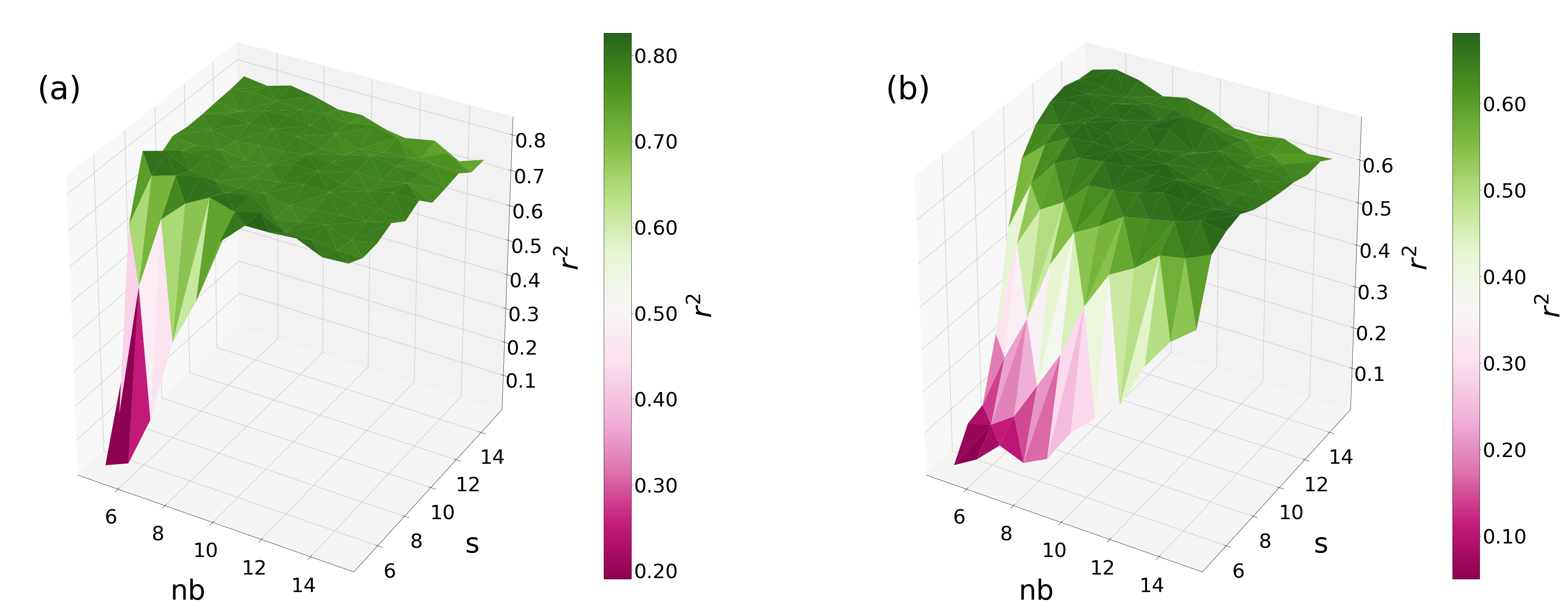}
\caption{R-squared values, $r^{2}$, dependency on the $nb$ number (x-axis) and $s$ size (y-axis) of sampled environments. Figure (a) shows results for the conference network and Figure (b) for the primary school network. Colours and z-axis code the obtained average $r^2$ score values for given $nb$ and $s$ parameter pairs computed over $10$ realisations. $\alpha$ was fixed to $0.5$; we set $d=20$ for Figure (a) and  $d=24$ for Figure (b) - see Figure \ref{fig:entropy_01}.}
\label{fig:param_context1}
\end{figure}

Next, we investigate how hyper parameters of the environment sampling may impact the prediction score on different real networks. Figure \ref{fig:param_context1} shows the $r^2$ scores computed for the conference and primary school networks with respect to the length $s$ and number $nb$ of environments sampled for each event. For these computations we fixed $\alpha=0.5$ and the embedding dimensions to their optimal values. According to Figure \ref{fig:param_context1} on the conference network, except for very small number or length of contexts we see an emerging plateau of $r^2$ values. This means that the environment size compensates for the number of environments (or vice versa) when we measure the embedding performance. In other terms, increasing the length of the environment has the same effect than increasing the number of environments on the $r^2$ score. For the primary school network we observe a similar but somewhat weaker compensation effect (Figure \ref{fig:param_context1}), while we observed the same behaviour even for the hospital and the high school networks (see SI). These results suggest that these two parameters are highly redundant, thus we can effectively reduce our parameter set by fixing both to a large enough value.

\begin{figure}[h!]
\centering
\includegraphics[width=0.9\linewidth]{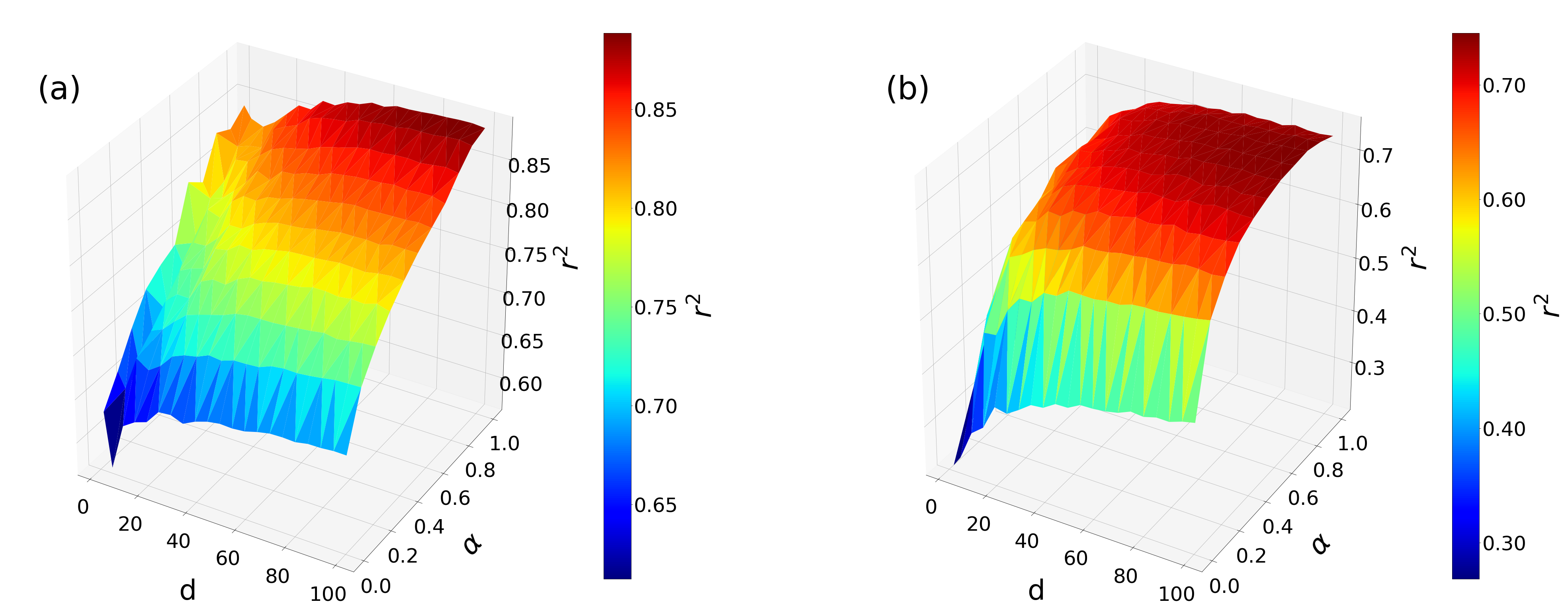}
\caption{R-squared values, $r^{2}$ , as the function of the $d$ number of embedding dimensions (x-axis) and $\alpha$ sampling balance (y-axis) parameters. Figure (a) shows results for the conference network and Figure (b) for the primary school network. Colours and z-axis code the obtained average $r^2$ score values for a given $d$ and $\alpha$ parameter pairs computed over $10$ realisations. Other hyper parameters were fixed to $nb=10$ and $s=10$.}
\label{fig:alpha_dim}
\end{figure}

In order to investigate the influence of the embedding dimension and the $\alpha$ sampling balance parameter, next we fix the environment parameters to $s=10$ and $nb=10$ based on our evaluation above. As shown in Figure \ref{fig:alpha_dim} (and in SI for other networks) both increasing the number of embedding dimensions and $\alpha$ lead to better performances in predicting the spreading outcome. As the function of the number of dimensions, each case reaches a plateau in accordance with our earlier results presented in Section~\ref{sec:entropy}. On the other hand we observe somewhat stronger dependencies on $\alpha$. While for the conference and the hospital networks the more one increases $\alpha$ the better the prediction gets, for the primary school and the high school networks the score reaches a plateau and become less sensitive to the change of $\alpha$ (see Figure \ref{fig:alpha_dim} and SI). If we consider lower values of $\alpha$, the similarity we capture between the event between adjacent events is mainly based on the co-occurrences, which are more relevant in school networks where participants might be active at the same time (e.g. in breaks between classes). This argument only moderately applies to a conference or hospital where simultaneous interactions typically happen in smaller groups or not at all. Higher values of $\alpha$ imbalance the sampling to contain more information about temporal paths, which actually indirectly codes co-occurrence frequencies as well. This gives the advantage to the model to learn both type of similarities and to predict the epidemic outcomes with higher precision.

\subsection{Comparison with other methods}
\label{sec:comparison}

Although our method seems to provide high absolute scores for the prediction of spreading outcome, there are a few other recently proposed dynamical network embedding methods, which can be used for the same task. Here we consider two of the most promising ones, the STWalk \cite{pandhre2018stwalk, stwalk}, and the Online-Node2vec embedding methods \cite{beres18on2v,online_n2v} to compare their predictive performances to weg2vec. Both methods are thought to build node embeddings for dynamic graphs using the Skip-Gram model, which introduces a significant difference to our event embedding method.

STWalk is designed to learn trajectory representations of nodes in temporal graphs by operating with two graph representations, a graph at a given time step and a graph from past time steps. It performs random walks respectively called space-walk and time-walk, to sample environments to input for the Skip-Gram embedding. The authors propose two variants of STWalk, different in the way the environment is built. In STWalk1, space-walk and time-walk are performed as part of a single step on a combined graph, while in STWalk2, space- and time-walks  are done separately.

Online-Node2vec is a node embedding method updating coordinates each time a new event appears in a temporal network. It also applies random walks to generate environments possibly using two strategies, the Temporal Walk algorithm and the Temporal Neighbourhood algorithm. In the Temporal Walk algorithm~\cite{beres2018temporal} a temporal path based centrality metric is used to capture similarity between nodes by projecting nodes on the same temporal path close to each other in the embedding. In the Temporal Neighbourhood algorithm\cite{fogaras2005scaling}, node similarity is inferred via a fingerprinting method, which projects nodes with similar neighbourhoods close to each other.

To compare the performance of the different methods, we test all of them on the four empirical networks introduced earlier. The environment parameter $nb$ and $s$ have been set to $10$ and $10$ for all cases to give them the same amount of information to learn and for a fair comparison of outcome. Further, we fix the balance parameter $\alpha$ to $0.5$. We then compute the average $r^2$ scores of simulated spreading outcomes as we vary the number of embedding dimensions. Since STWalk and Online-Node2vec use only the past and the present as basis for the nodes environment, we run the simulation for our methods using only the predecessors for each event as well (see Section \ref{sec:context}). As previously we estimate the epidemic size by using the coordinates of the actual embedding in a linear regression model (see Section \ref{sec:spreading}).

\begin{figure}[h!]
\centering
\includegraphics[scale=0.11]{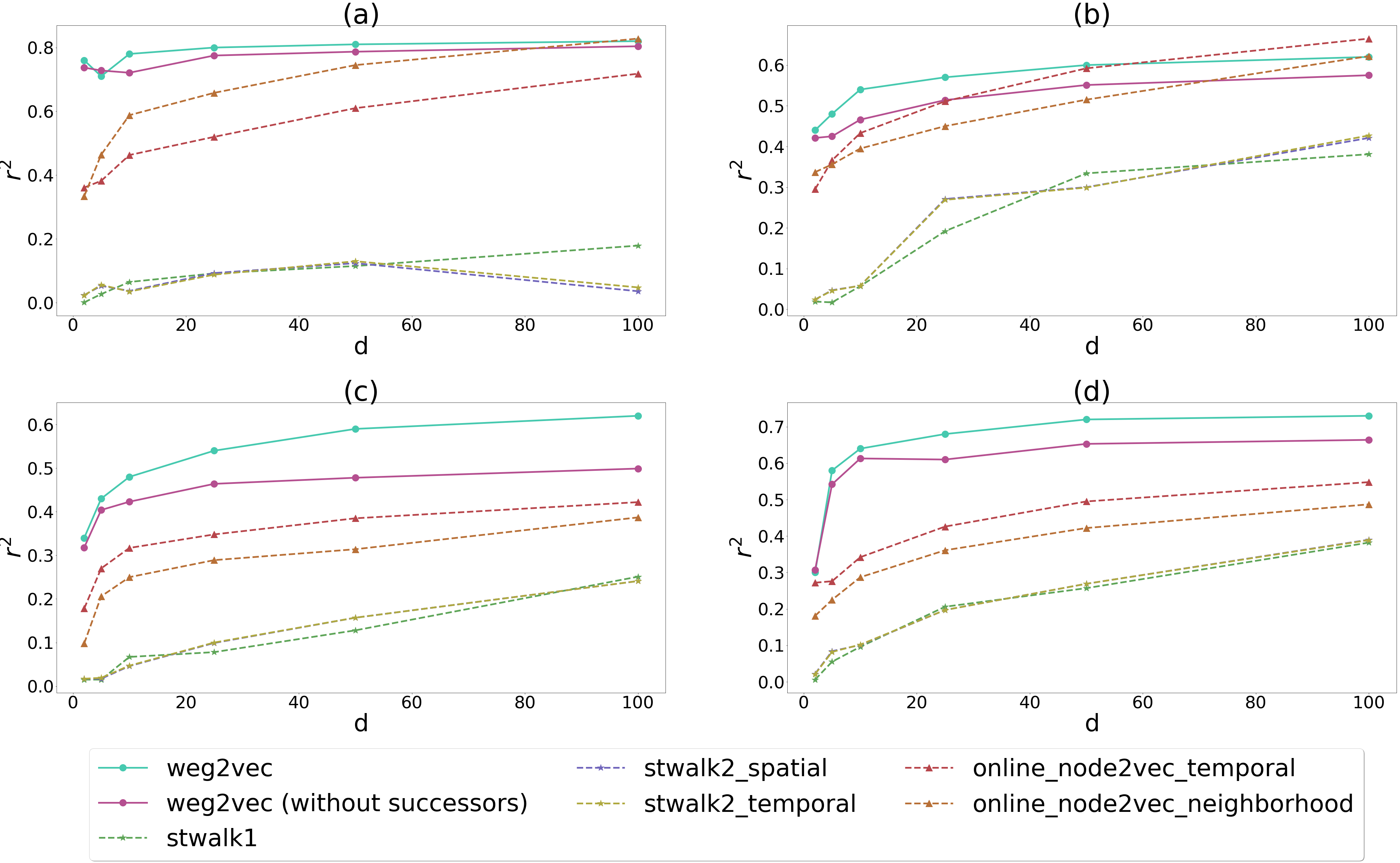}
\caption{Comparison of STWalk, Online-Node2vec and our embedding methods in predicting spreading outcomes on empirical networks in different settings as (a) conference, (b) hospital, (c) high school, and (d) primary school. Results shown are $r^2$ scored obtained from linear regression on coordinates in embedding spaces with various dimensions computed for each method and empirical temporal networks.}
\label{fig:compare_original}
\end{figure}

According to the results in Figure \ref{fig:compare_original}, our method outperforms all the other methods on any of the networks for a broad range of dimensions. The performance improves if we also consider the successors and not only the predecessors in building the environment, as expected. The exception is the hospital network, where our method gets lower scores with respect to Online-Node2vec for dimensions $50$ or larger. In general, the difference in the scores can be explained due to the advantage of event embedding instead of node embedding. Indeed if we are looking at epidemic spreading mediated by temporal interactions, it becomes more natural to work with events . In the case of STWalk, the lower scores can be partly explained by the selection of the environments that are allowed to included higher-order correlations among nodes. This more complex information coded in the environments can appear less relevant or noise for the learning task in this case. In case of Online-Node2vec, the relative under-performance can be due to the fact that information of the temporal and neighbourhood information are considered separately instead. Missing to join these two aspects, can lead to limited information and limited prediction capacities.

\section{Discussion}
\label{sec:conclusions}

Embedding of networks has recently drawn a lot of attention as it both provides lower dimensional representations of networks and proves to be efficient to resolve task such as link prediction, node classification or anomaly detection. Here we proposed an embedding technique of temporal networks, a domain still little explored. Indeed most low dimensional representations of networks introduced are for static networks. Moreover instead of generating node embedding, we are creating link embedding which we show to be much more efficient when we try to solve a task linked to spreading process compared to node embedding techniques.

The embedding method we introduced has the advantage to be very simple, it relies on the sampling of neighbourhood on a higher order static representation of the temporal networks and the use of the Skip-Gram model, largely developed for word embedding. The neighbourhood sampling was built so that it takes into account both the notions of causal temporal paths and co-occurrences, meaning that events that are on the same temporal paths and that tend to co-occur are projected close in space. We have shown that an embedding based on this neighbourhood sampling is particularly efficient to provide compact representations of temporal networks that retain essential features of the networks such as the time ordering and the organisation of networks in mesoscale structures. Here, we also provide a way to choose a relevant dimension for the embedding. Along with this, we show that the learned representations retain enough information of the original network to get a relevant estimate of the outcome of spreading process. Interestingly, this observation remains true even if the sampling strategy uses only information from the past for each event. This means that the technique introduced can be also used as an online method taking into account temporal events on the run. Moreover, tuning the neighbourhood sampling, i.e. playing on the trade-off between the inclusion of the concepts of causal temporal paths and co-occurrences, to get the best performance for the prediction of the outcome of the spreading process can be used to detect the relevant properties of the original network for the spreading process. 

For future works, it would worth exploring other sampling strategies that decouples the purely structural properties, i.e. the presence of the communities in the aggregated network, from the temporal properties. Another important follow-up of this work would be the application of this embedding technique to solve questions such as the detection of key events in misinformation spreading.

\section{Methods}
\subsection{Entropy Computation}
\label{sec:entropy_method}

To measure the entropy over the distributions of the euclidean distances between pairs of event coordinates in each of the temporal network embedding, we build 10 embeddings for $50$ different dimensions (from 2 to 100 at step 2), setting $\alpha = 0.5$.  The hyper parameters $nb$ and $s$ were set to $10$ and $10$ respectively. For each embedding, we then divide our dataset in $10$ consecutive samples of 1000 consecutive events each, both to avoid to compute all the pairwise distances (which would be very costly computationally)  and at the same time have a representative set of events.

We compute the euclidean distance of each pair of events in the samples for each of the $10$ different embeddings. We then bin the distances into $k=10$ bins ranging from the global maximum and the global minimum values over all the possible dimension and realisations of the embedding for the same network, and measure the entropy over these sets of distances as
\begin{equation}
    H = \sum_k {p_{k} \log p_{k}},
\label{eq:entropy}
\end{equation}
where $p_{k}$ represents the probability associated with the $k_{th}$ bin. In Figure \ref{fig:entropy_01} (see also SI) the blue curve represents the entropy values with respect to the number of embedding dimensions, averaged on the 10 samples as described above and the shaded surrounding area shows the variance among the 10 samples. The vertical dash line corresponds to the dimension at which the embedding stabilises. To determine this point we looked for the best fit of a horizontal line on the average entropy curve and took the value of the first interception of the curve with its fit.

\subsection{Spreading Process}
\label{sec:spreading}

The simplest epidemic models are based on the assumption that the population can be divided into compartments, each representing a phase of the disease \cite{anderson1992infectious, bailey1975mathematical, daley2000epidemic, diekmann2000mathematical, barrat2008dynamical}. The one we used for our analysis, the Susceptible-Infected model, which foresees that once a healthy node (in state $S$) is exposed to the infection, it will become infected (state $I$) with a given rate $\beta$ and will never return to the original healthy state. In our specific case, the $SI$ model has been implemented in such a way that $\beta=1$, which defines a a deterministic process from the spreading point of view.

Taking each event as the starting of the epidemic, we simulate the epidemic spreading on the temporal network and we assign the final epidemic size to each seed event. We build a dataset containing embedding coordinates of each event, the associated epidemic size that will be the target to be predicted, the square of each coordinate and the euclidean distance from each event in the network and the first event in time, that will be used as regressors. In other terms, we assume a linear relationship between the epidemic size ($y$) and the embedding coordinates ($\vec{x}$) defined by the so-called \textit{regression equation}
\begin{equation}
    y = \beta_0 + \sum_{i=1}^r {\beta_i x_i}
\label{eq:linear_regression}
\end{equation}
where  $x_1,...,x_r$ are the predictors (the embedding coordinates, their square and the euclidean distance) and $\beta_0,...,\beta_r$ are the regression coefficients.   For training we operate with a $10$-fold cross-validation, i.e. we first randomly partitioned the original sample into $10$ equal sized sub-samples and retain a single sub-sample as the validation data for testing the model, while using the remaining $9$ sub-samples for the actual training. We repeat this process $10$ times to train the embedding to best learn the coordinates of each event in the network.  We use the coefficient of determination, denoted as $r^2$, to understand which amount of variation in $y$ can be explained by the dependence on $\vec{x}$ using the particular regression model. Larger $r^2$ indicates a better fit and means that the model can better explain the variation of the output with different inputs.

\subsection{Tensor Factorisation for mesoscale structure extraction}
\label{sec:tensor_decomposition}

A temporal network can be fully described by a time‐dependent adjacency matrix, where its entries are either one or zero depending on the presence or absence of interactions of pair of nodes at a given time. This matrix can be seen as a three-way tensor, whose size is $n\times n \times t$ ($n$ indicates the number of nodes and $t$ the number of snapshots in the temporal network). This tensor can then be factorised into a sum of $r$ rank-one tensors, i.e. number of mesoscale structures we search into the temporal network. Using this technique we can group events into  mesoscale structures. Indeed with the decomposition, we now have a value for each link for each time for each rank-$1$ tensor, we assign a mesoscale structure to each event based on these values. Basically the mesoscale structure assigned to each event is the one for which the corresponding link at the corresponding time has the highest value. To find the optimal number of mesoscale structures, we used the core consistency metric \cite{bro2003new}. It is based on scrutinising the \textit{appropriateness} of the structural model based on the data and the estimated parameters of gradually augmented models. A model is called \textit{appropriate} if adding other combinations of the same components does not improve the fit considerably. In practice, we operate different tensor decompositions for different value of the rank (ranging from 2 to 20, for all the networks) in order to estimate the best value for it.

\section*{\textbf{Supplementary Information}}

\subsection*{Temporal network data}
\label{sec:details_nets}
In order to demonstrate the performance of the weg2vec method for the embedding of events in real networks we used temporal network datasets from the SocioPatterns project~\cite{cattuto2010dynamics}. We concentrated on four different settings, in a conference, hospital, primary school and high school, where we could expect particularly different interaction dynamics and in turn different final outcome of the simulated spreading process. General details of the datasets are summarised in in Table \ref{tab:net_features_appendix}. While we demonstrated all results in the main text using the conference and primary school networks, we repeated all calculations on each of them with results presented in the Supplementary Information. 

\begin{table}[ht]
\begin{center}
\begin{tabular}{|l|c|c|c|}
\hline
\textbf{Network}&\textbf{Nodes}&\textbf{Events}&\textbf{Temporal Interval}\\ \hline
conference&113&10457& $\sim$ 2.5 days\\
\hline
hospital&75&13650&$\sim$ 4 days \\
\hline
high school&327&36015&$\sim$ 2 days $\ast$ \\
\hline
primary school&236&35921&$\sim$ 8 hours $\ast$ \\
\hline
\end{tabular}
\caption{Main features of the empirical temporal networks we analyze in our paper.}
\label{tab:net_features_appendix}
\end{center}
\vspace{-5mm}
\end{table}

$\ast$ We concentrated only on certain periods of the high school and primary school networks in order to have a consistent number of events across all datasets. For the high school, we selected from time stamp 1385982020 to time stamp 1386156680, while for the primary school we selected from time stamp 31220 to time stamp 60020.

\subsection*{Embedding of temporal networks}

\begin{figure}[!h]
\centering
\includegraphics[width=0.9\linewidth]{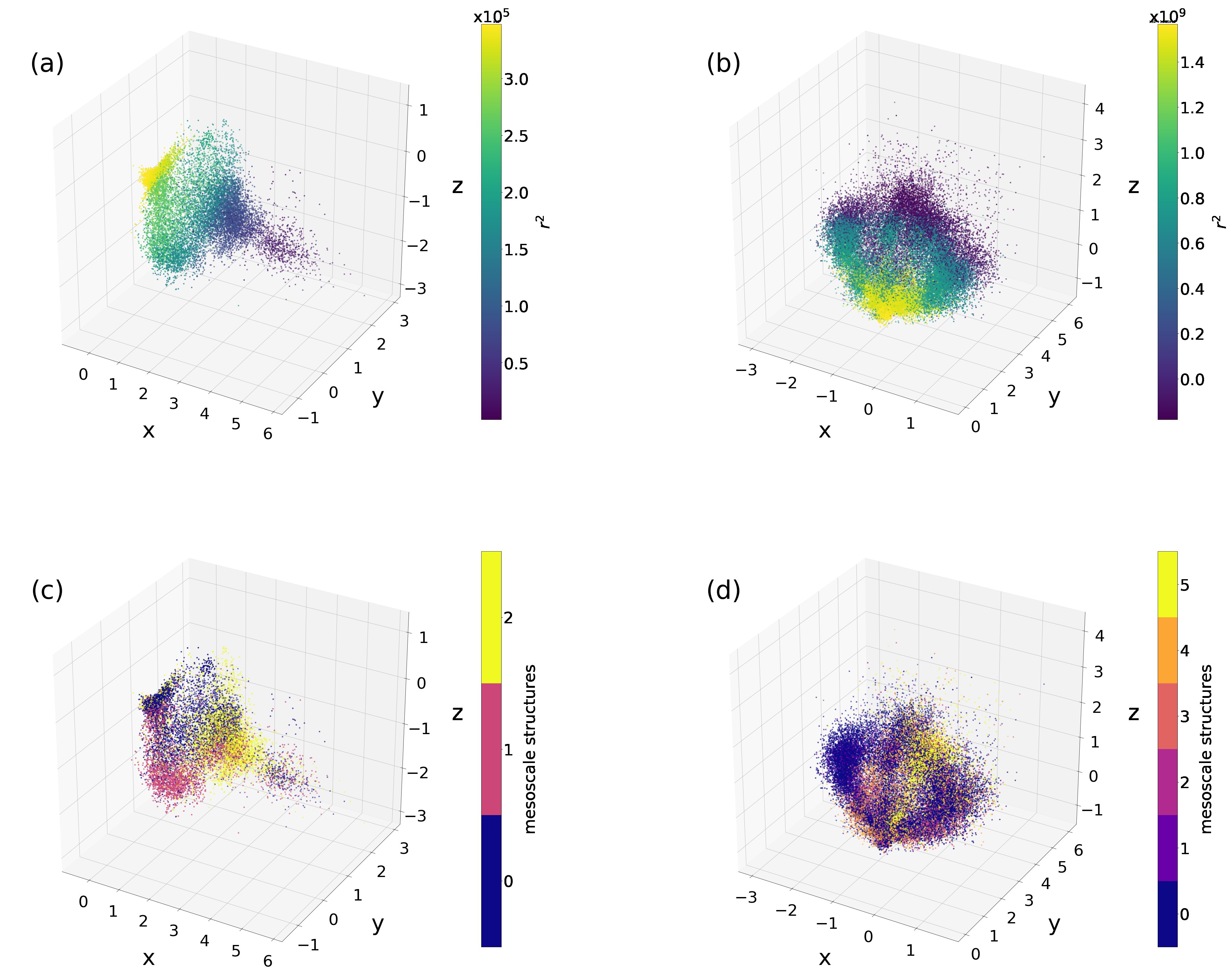}
\caption{3-dimensional embeddings of the hospital and high school networks. Colours in panels (a-b) shows the time at which the event occurs and in panels (c-d) it indicates mesoscale structure membership detected by tensor factorisation method with $3$ and $6$ groups (respectively). Hyper parameters were set to $\alpha=0.5$, $nb=10$ and $s=10$.}
\label{fig:3d_emb_appendix}
\end{figure}

As we demonstrated in the main text, the weg2vec method successfully captures multiple types of correlations present in a real network. This is demonstrated in Figure \ref{fig:3d_emb_appendix} for the hospital and high school networks (and for other networks in the main text). These figures show a $3$-dimensional embedding of temporal network events, represented each as a point in the vector space. On Figures~\ref{fig:3d_emb_appendix} (a) and (b), colours represent the time at which event occurs in the temporal network. The gradient change of colour indicates that the embedding captures correctly the time ordering of the events. Figure \ref{fig:3d_emb_appendix} (c) and (d) show the same $3$-dimensional representation but with colours representing the membership of events in mesoscale structures detected by tensor factorisation techniques (for more details on this method see Material in the main text). The grouping of events of the same colour suggests that the embedding is able to capture some of these mesoscale structures.

\subsection*{Parameter dependence}

\label{sec:selection_params}
\begin{figure}[!h]
\centering
\includegraphics[width=0.9\linewidth]{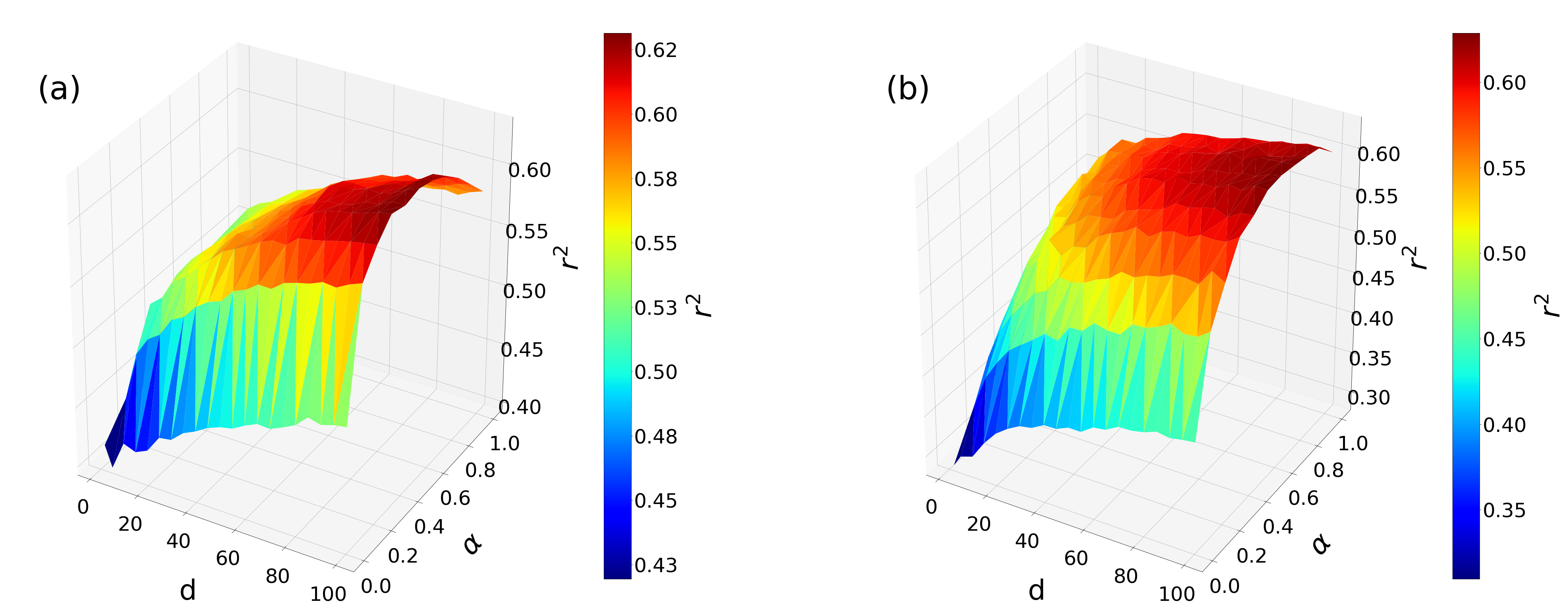}
\caption{The $r^{2}$ correlation score dependency on the $d$ number of dimensions and $\alpha$ sampling balance parameters. Results are shown for the (a) hospital and (b) high school networks. Colours and z-axis code the obtained average $r^2$ score values for a given $d$ and $\alpha$ parameter pairs computed over $10$ realisations. Other parameters were fixed to $nb=10$ and $s=10$.}
\label{fig:alpha_dim_appendix}
\end{figure}

Similar to the results presented in the main text, here we repeated the analysis to screen the parameter dependence of the prediction of spreading outcomes for the hospital and high school networks. In Figure \ref{fig:alpha_dim_appendix} we fixed the dimensions to the detected optimal values ($d=14$ for the hospital data and $d=26$ for the high school networks) and also set $\alpha$ to $0.5$. At the same time we varied the environment hyper-parameters $nb$ and $s$. Results are very similar to the reported ones in the main text, showing that these parameters are strongly coupled. As we increase them a plateau of $r^2$ emerges where the prediction become invariant of these parameters beyond statistical fluctuations. Consequently, choosing a large enough value for both of these hyper-parameters would be optimal for the prediction task.
 
\begin{figure}[!h]
\centering
\includegraphics[width=0.9\linewidth]{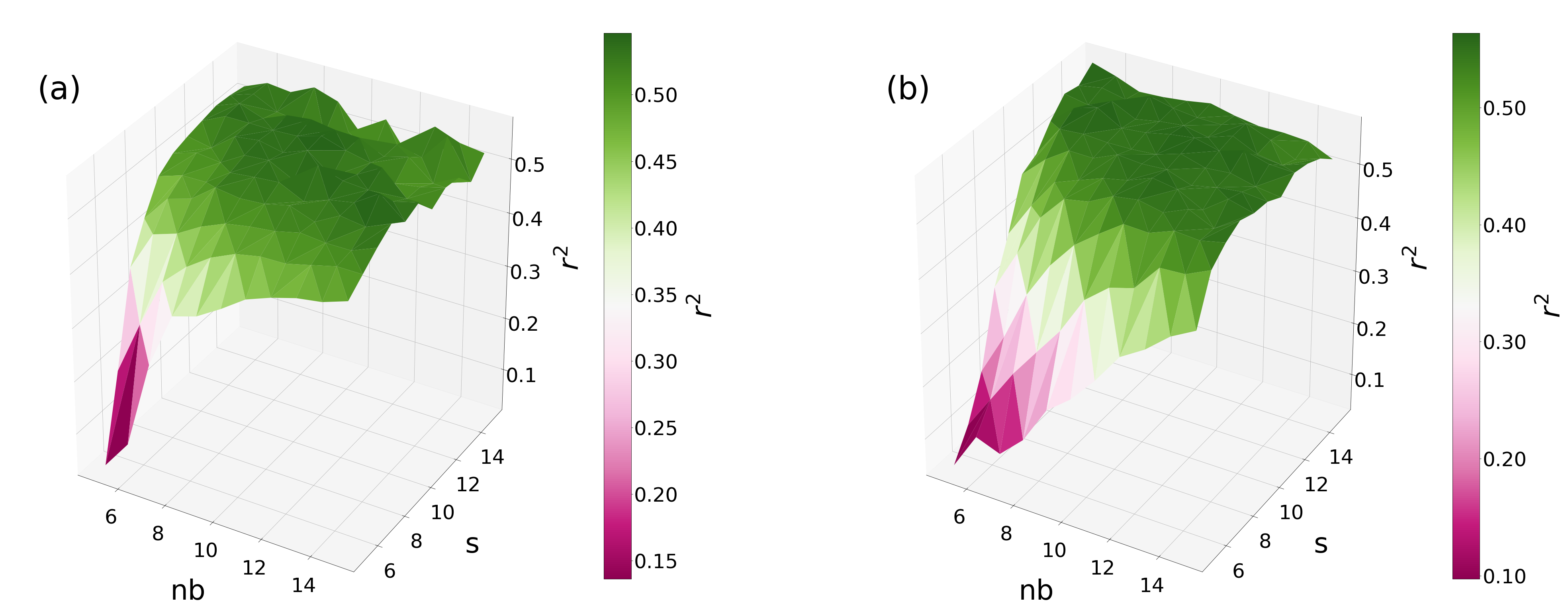}
\caption{The $r^{2}$ correlation score dependency on the $nb$ number and $s$ size of sampled environments. Results are shown for (a) hospital and (b) high school networks. Colours and z-axis code the  average $r^2$ score values for given $nb$ and $s$ parameter pairs computed over $10$ realisations. Other parameters were fixed to $\alpha=0.5$ and (a) $d=14$ and (b) $d=26$.}
\label{fig:param_context_appendix}
\end{figure}

On the contrary, by fixing the environment parameters $nb$ and $s$ to 10 and varying $\alpha$ and the number of embedding dimensions, we see stronger parameter dependencies. Similarly, as we increase the $\alpha$ parameter the performance improves but the system shows stronger response. While the performance gets better by increased dimensions, after a given value the improvement is marginal or may even decrease (in case Figure~\ref{fig:alpha_dim_appendix}(a)). In terms of dimensions the scaling of $r^2$ initially shows rapid improvements of the prediction task but after a certain number of dimensions the improvement is only marginal, indicating an optimal dimension number for training, in agreement with our entropy analysis.

\subsection*{Entropy measures}

\begin{figure}[!h]
\centering
\includegraphics[width=0.9\linewidth]{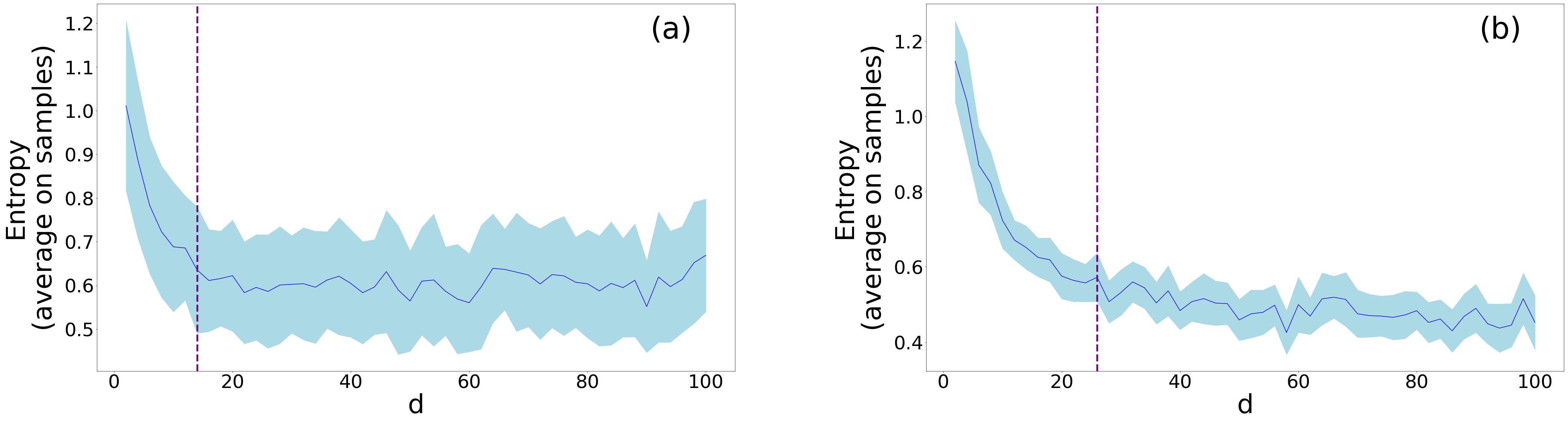}
\caption{Entropy values with respect to the $d$ number of embedding dimension for the (a) hospital and (b) high school networks, with $\alpha = 0.5$, $s=nb=10$. The dash line indicates the optimal embedding dimension in which stability is reached. The blue line and the shaded area represent respectively the average and the variance among the samples we used for the analysis.}
\label{fig:entropy_01_appendix}
\end{figure}

We performed the same entropy analysis on the hospital and high school networks what was presented for other datasets in the main text. Similarly, we measured the entropy of the variance of the relative distance between embedded events as the function of the number of embedding dimensions. Qualitatively we found the same results (see Figure~\ref{fig:entropy_01_appendix}) as reported in the main text, that the entropy measure becomes invariant of the number of dimensions after an optimal value and involving further dimensions would not increase significantly the quality of the prediction but would only introduce redundancies in the representation.

\subsection*{Epidemic size distribution of original networks and randomized models}

\begin{figure}[h!]
\centering
\includegraphics[width=0.7\linewidth]{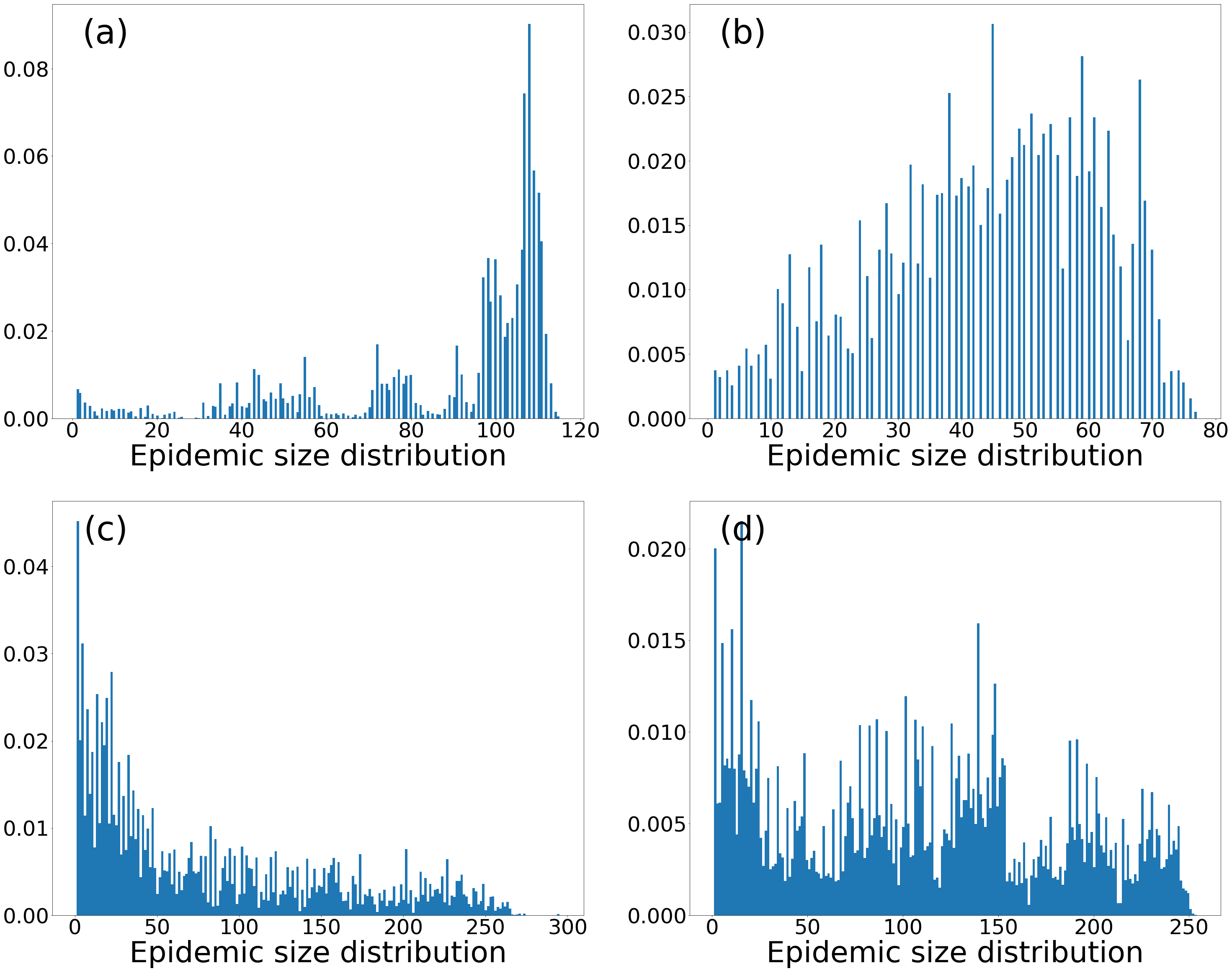}
\label{fig:sub1_ep}
\caption{Epidemic size distribution of the (a) conference, (b) hospital, (c) high school, and (d) primary school original temporal networks.}
\label{fig:ep_size_dib_original}
\end{figure}

While in the main text we only discussed the correlations between the simulated and predicted epidemic sizes, we never looked at how simulated epidemic sizes are varying over several realisations. Here we look at these distributions measured for simulated epidemic processes seeded from every event in the empirical networks. As shown in Figure~\ref{fig:ep_size_dib_original} for each dataset, we find that these distributions vary in different ways for different networks. While they are concentrated for the (a) conference and (c) high school datasets, they are more homogeneously distributed for the (b) hospital and (d) primary school data.

\begin{figure}[h!]
\centering
\includegraphics[width=0.7\linewidth]{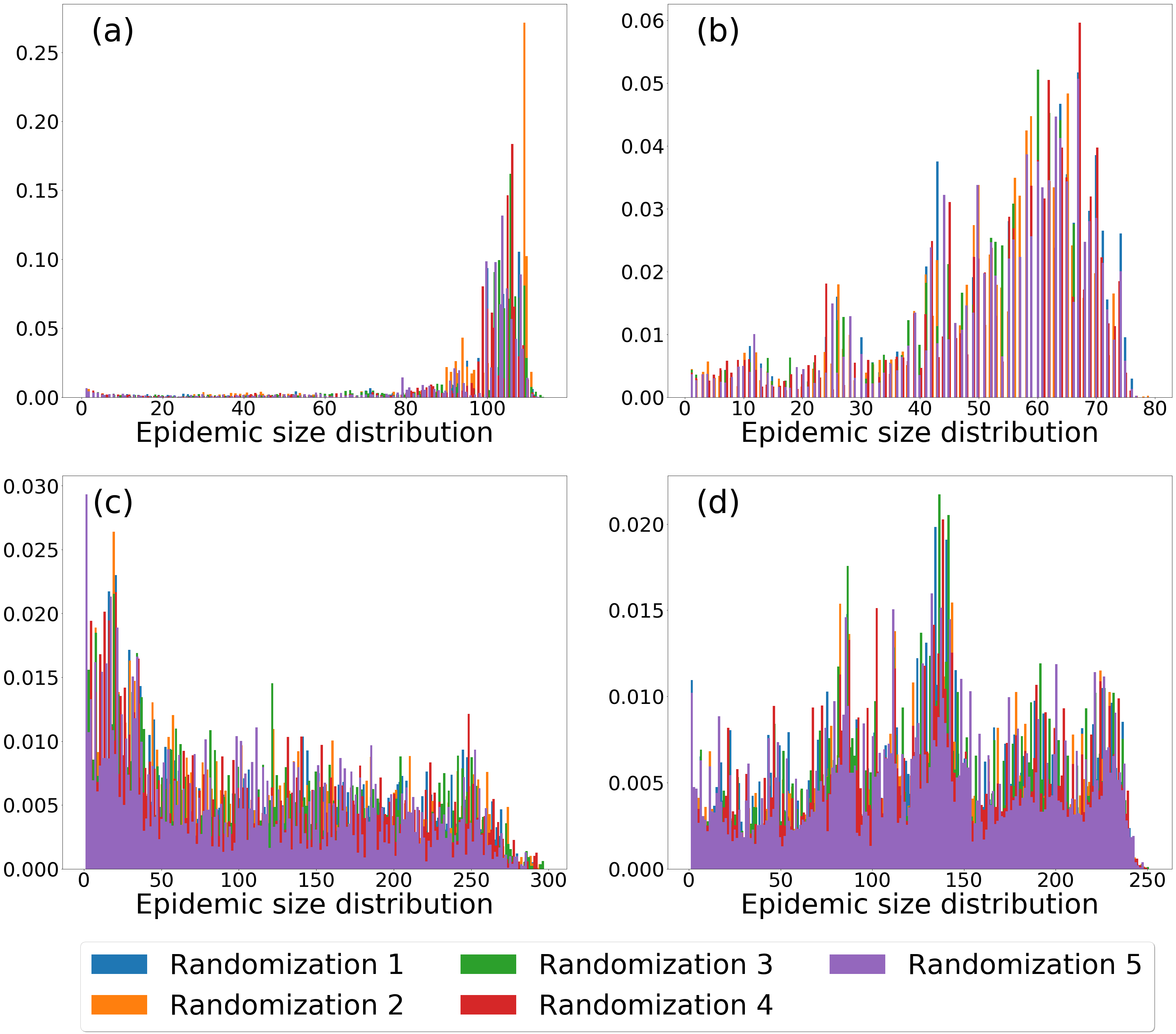}
\caption{Epidemic size distributions on \emph{snapshot shuffled} RRM networks for (a) conference, (b) hospital, (c) high school, and (d) primary school networks. Colours assign different random realisation of the actual network model.}
\label{fig:ep_size_dib_t}
\end{figure}

\begin{figure}[h!]
\centering
\includegraphics[width=0.7\linewidth]{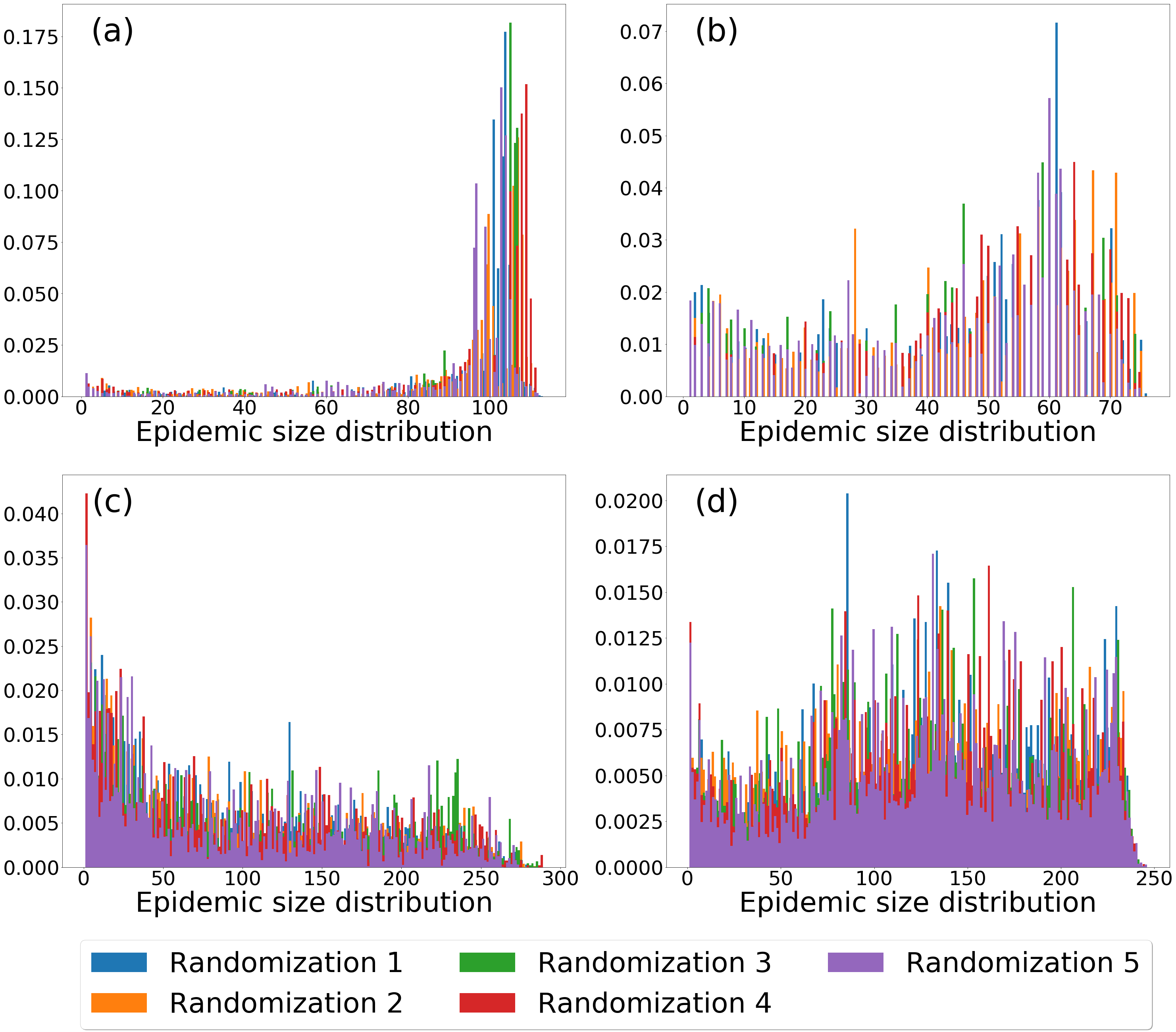}
\caption{Epidemic size distributions on \emph{timeline shuffled} RRM networks for (a) conference, (b) hospital, (c) high school, and (d) primary school networks. Colours assign different random realisation of the actual network model.}
\label{fig:ep_size_dib_l}
\end{figure}

To further understand the effects of the different RRMs on the epidemic outcome, for each models we generated five different randomised network realisation and simulated the spreading process starting from each event in the networks to obtain the distribution of final epidemic sizes. Results shown in Figure~\ref{fig:ep_size_dib_t} for the snapshot shuffling RRM, and in Figure~\ref{fig:ep_size_dib_l} for the timeline shuffling RRM indicate that these models do not alter considerably the final outcomes of the the spreading processes as the obtained distributions are remarkably similar to the ones measured on the empirical network (see Figure~\ref{fig:ep_size_dib_original}). On the other hand, for the link shuffling RRM the distributions become surprisingly narrow around relatively large values. This suggests that the observed variance of the epidemic outcomes is largely due to structural correlations and potentially due to temporal correlations between events on adjacent links. At the same time, this explains why this method performs the best, sometimes even better than the real network, in the prediction task. Predicting a narrow outcome of a process is a considerably easier task than predicting a process with an outcome of high variance.

\begin{figure}[h!]
\centering
\includegraphics[width=0.7\linewidth]{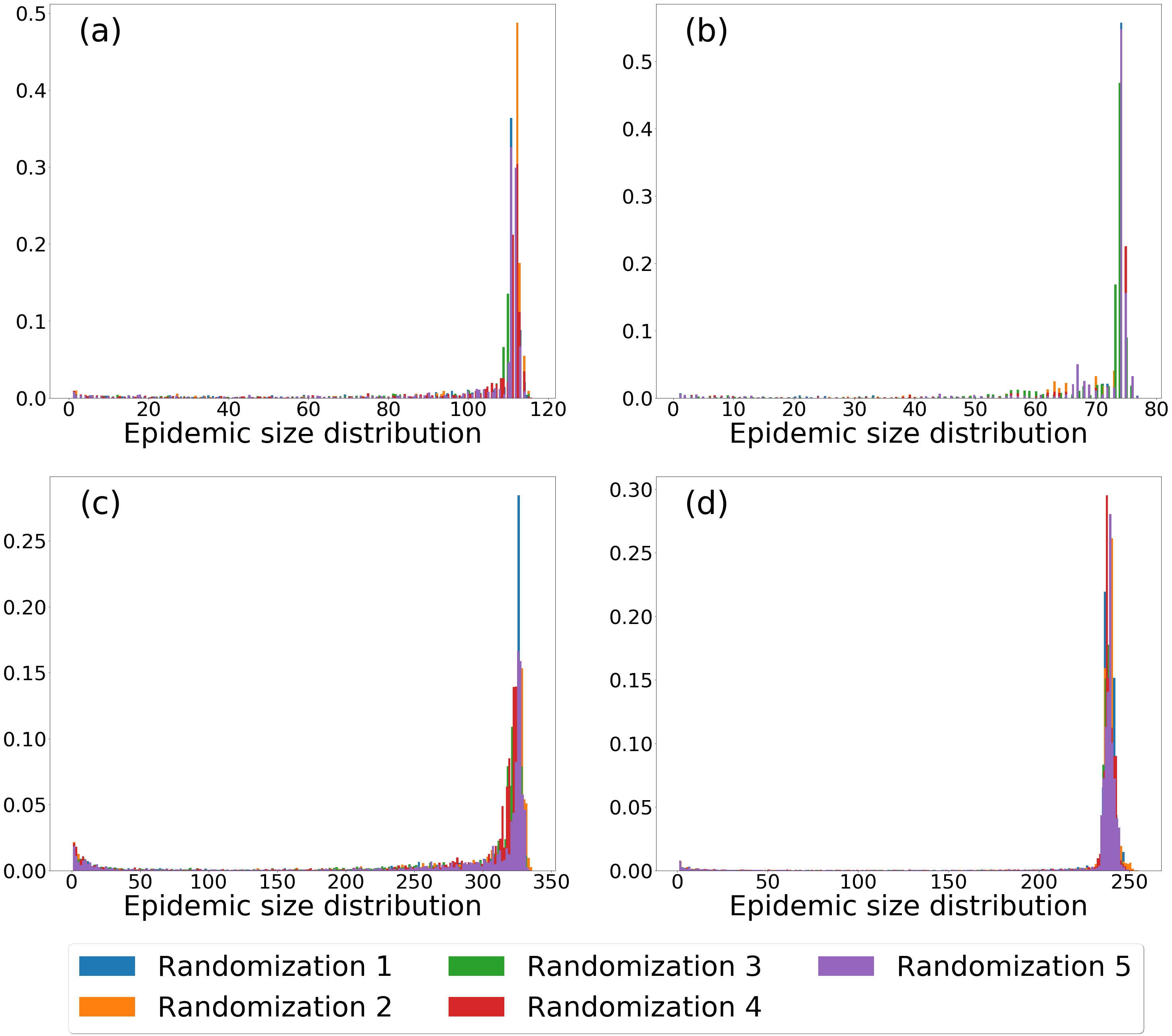}
\caption{Epidemic size distributions on \emph{link shuffled} RRM networks for (a) conference, (b) hospital, (c) high school, and (d) primary school networks. Colours assign different random realisation of the actual network model.}
\label{fig:ep_size_dib_n}
\end{figure}

\end{document}